\documentclass[twocolumn]{aastex631}

\shorttitle{Spin of MAXI J1727-203}

\graphicspath{{./}{fig/}}
\begin{document}

\title{Estimating the Black Hole Spin for the X-Ray Binary MAXI J1727-203 Based on Insight-HXMT}

\author[0000-0002-4858-3001]{Haifan Zhu}
\affiliation{Department of Astronomy, School of Physics and Technology, Wuhan University, Wuhan 430072, China}
\author[0000-0003-3901-8403]{Wei Wang}
\altaffiliation{Email address: wangwei2017@whu.edu.cn}
\affiliation{Department of Astronomy, School of Physics and Technology, Wuhan University, Wuhan 430072, China}

\begin{abstract}

We constrain the spin of the black hole (BH) candidate MAXI J1727-203 using Insight-HXMT data. Due to limited HXMT observations covering only part of the outburst, NICER data were used to analyze the full outburst's state transitions, we identified two of three HXMT observations in the high soft state and applied the continuum-fitting method to measure the spin. Based on previous estimates and continuum spectral fittings, we explored the parameter space and found that the best-fitting values were $(D, i, M) \approx (6\ \text{kpc}, 30^\circ, 12 M_{\odot})$. We also tested the variation of these parameters using Monte Carlo simulations, sampling over 3000 sets within the parameter ranges: $5.9 \text{kpc}< D<7  \text{kpc}$, $24^\circ<i< 35^\circ$, and $10 M_{\odot}<M<14 M_{\odot}$, yielding a spin measurement of $a=0.34_{-0.19}^{+0.15}$ (1$\sigma$). 
In addition, we analyzed NuSTAR data in low hard state and found a good fit with the {\tt tbabs*(diskbb+powerlaw)} model, with no significant iron line features observed in the residuals, then the previous reflection model results suggesting an extremely high spin would over-estimate the BH spin.
\end{abstract}

\keywords{High energy astrophysics (739); Black hole physics (159); Stellar mass black holes (1611); Stellar accretion disks (1579); X-ray transient sources (1852); X-ray sources (1822); X-ray binary stars (1811)}

\section{Introduction}
\label{sec:intro}

Black Hole X-ray Binaries (BHXRBs) are binary systems in which a black hole accretes matter from a companion star through Roche lobe overflow. These systems spend most of their time in a quiescent state, with occasional outbursts that can last from weeks to months, during which the X-ray intensity increases by several orders of magnitude compared to the quiescent state. During an outburst, the BHXRBs evolves through various spectral states \citep{remillard2006x,belloni2010states}. The various spectral states of a BHXB can be broadly categorized as the low hard state (LHS),
intermediate state (IMS), and high soft state (HSS). Its evolutionary trajectory starts from  LHS, progresses through  IMS, then to HSS, and finally returns to LHS, forming a "q"-shaped track in the hardness-intensity diagram (HID, \citealt{fender2004grs}).

The black hole spin is a fundamental parameter that provides insights into its formation and evolution while influencing a variety of astrophysical phenomena. The spin parameter of accreting stellar-mass black holes is commonly measured using two established methods: (1) the continuum-fitting method, which analyzes the thermal emission profile of the accretion disk \citep{zhang1997black}, and (2) the reflection-fitting method, which focuses on the red wing of the relativistically broadened and asymmetric Fe $\rm K\alpha$ line \citep{fabian1989x,reynolds2003fluorescent}.

Both methods use a dimensionless parameter, $a_{\star}$, to represent the black hole spin, defined as $a_{\star} \equiv \frac{a}{M} = \frac{cJ}{GM^2}$, 
where $a_{\star}$ represents the black hole spin, $M$ and $J$ denote the black hole's mass and angular momentum, respectively \citep{kerr1963gravitational}. Both methods are based on the fundamental assumption that the inner edge of the accretion disk extends to the innermost stable circular orbit (ISCO). The key to measuring the spin lies in estimating the radius of the inner disk, $r_{\text{in}}$, defined as  
$r_{\text{in}} \equiv \frac{cR_{\text{in}}}{GM}$. It is assumed that $R_{\text{in}}$ corresponds to the radius of the ISCO, $R_{\text{ISCO}}$. Since  $R_{\text{ISCO}}/(GM/c^2)$ maps the spin parameter $a_\star$ in a simple and monotonic manner, determining $R_{\text{in}}$ (assuming $R_{\text{in}} = R_{\text{ISCO}}$) directly enables the derivation of the black hole spin $a_\star$ \citep{reynolds2008broad,noble2009direct,penna2010simulations,kulkarni2011measuring}. 

In the continuum-fitting method, the inner disk radius, $r_{\text{in}}$, is determined by fitting the X-ray thermal continuum from the accretion disk to the Novikov-Thorne thin disk model \citep{novikov1973black}. As a non-relativistic approximation, the disk luminosity can be expressed as  
$L \approx 2\pi D^{2} F (\cos i)^{-1} \approx 4\pi R_{\text{ISCO}}^2 T_{\text{eff}}^4,$
where $F$, $T_{\text{eff}}$, $D$, and $i$ represent the X-ray flux, effective temperature, distance to the source, and disk inclination angle, respectively. This leads to the relationship  $R_{\text{ISCO}} \approx F/(2 T_{\text{eff}}^4 (\cos i)^{-1} D^2 M^{-2})$.  
Therefore, three key dynamical parameters—the black hole mass ($M$), disk inclination ($i$), and source distance ($D$)—are essential for accurately estimating the spin. The spin parameter $a_\star$ is inversely proportional to $R_{\text{ISCO}}/M$, implying that a larger black hole mass results in a higher spin. Conversely, an increase in either the inclination angle  or the source distance  increases the value of $R_{\text{ISCO}}/M$, thereby decreasing the estimated spin parameter \citep{bardeen1972rotating,zhao2021estimating}. In addition, uncertainties in the black hole mass $M$, the source distance $D$, and the inclination angle $i$ of the accretion disk can also impact the measurement. These factors represent the main sources of systematic uncertainty in the continuum-fitting method. 

MAXI J1727-203 was discovered by the MAXI/GSC nova alert system on 2018 June 5 as an uncatalogued X-ray transient located at (RA, Dec) = $(261.971^\circ, -20.389^\circ)$ (J2000) with a 90\% confidence elliptical error region of $0.33^\circ \times 0.28^\circ$ \citep{yoneyama2018maxi}. 
\cite{ludlam2018nicer} and \cite{kennea2018maxi} reported same-day observations of MAXI J1727-203 using the Neutron star Interior Composition Explorer (NICER) and the Neil Gehrels Swift Observatory, respectively.
Energy spectral softening observed by MAXI/GSC, combined with the detection of an ultrasoft component with a low disk temperature and large inner disk radius, strongly suggests that MAXI J1727-203 is a black hole binary \citep{negoro2018maxi}. Then a soft to hard transition was observed with Swift/XRT \citep{tomsick2018maxi}. 

\cite{alabarta2020x} presented an X-ray study of MAXI J1727-203's 2018 outburst using NICER data. The source exhibited soft and hard spectral components, evolving through HSS, IMS, and LHS. The disk component was detected throughout the outburst, with temperatures dropping from $\sim 0.4$ keV to $\sim 0.1$ keV. The power spectrum showed broad-band noise up to 20 Hz, with no QPOs. The fractional rms increased with energy, except in the hard state, where it remained constant. The spectral and timing evolution suggest the system hosts a black hole. \cite{wang2022multi} also studied the X-ray spectral and timing evolution of MAXI J1727 - 203 using NICER and MAXI/GSC data, and observed a transition from the IMS to the HSS, and back to the LHS. In the HSS, the innermost radius remained constant. Assuming a Schwarzschild black hole, the mass was estimated to be $M \geq 11.5 M\odot$ and $D \geq 5.9$ kpc, based on the multi-colour disk model using X-ray, optical and near-infrared multi-wavelength data \citep{wang2022multi}. 

With the observations from Insight-HXMT and NICER, we try to constrain the spin of the black hole in MAXI J1727-203 using the continuum-fitting method. Section \ref{obs} describes the observations and data reduction methods. Section \ref{RESULTS} provides the fitting results. Sections ~\ref{DISCUSSION} and \ref{conclusion} present the discussions and conclusions, respectively.

\section{Observations and Data Reduction}
\label{obs}
\subsection{Insight-HXMT}
 Insight-HXMT is equipped with three distinct types of detectors,  each designed for specific energy ranges. The High Energy (HE) detectors operate within the range of 20.0 to 250.0 keV \citep{liu2020High}. The Medium Energy (ME) detectors function between 5.0 and 30.0 keV \citep{cao2020medium}. Lastly, the Low Energy (LE) detectors cover a range from 1.0 to 15.0 keV \citep{chen2020low}. The effective areas for these detectors are 5100 $\rm cm^2$, 952 $\rm cm^2$, and 384 $\rm cm^2$, respectively.
\begin{table*}
\centering
\caption{Information of Insight-HXMT observations on MAXI J1727-203 }
 \setlength{\tabcolsep}{4.5pt}
\begin{tabular}{ccccccc}
 \hline\noalign{\smallskip}
ObsID &ExpoID & MJD&Obs data & LE exposure & ME exposure& HE exposure \\
 & &&(yyyy-mm-dd) & s &s&s\\
\hline
P0114758001 &P011475800101& 58277.229&2018-06-08 & 100.747& 308.206 &864.204  \\
P0114758001&P011475800102& 58277.377&2018-06-08 & 1539.141  & 2279.616&1109.173    \\
P0114758002 &P011475800201& 58279.681 & 2018-06-10 &   2692.251&2928.097&1330.825  \\
P0114758002 &P011475800202& 58279.846& 2018-06-10 &  1077.301 &1033.546&1440.614 \\
P0114758003 &P011475800301& 58282.729& 2018-06-13 &   2513.701&2729.343& 3137.429 \\
\noalign{\smallskip}\hline
\label{tab1}
\end{tabular}

\end{table*}

The Insight-HXMT observed MAXI J1727-203 three times, and we listed the parameters of each observation in Table ~\ref{tab1}.
Data extraction and analysis were carried out using Version 2.06 of the \textit{Insight-HXMT Data Analysis Software} (HXMTDAS)\footnote{\url{http://hxmten.ihep.ac.cn/software.jhtml}}. We processed and filtered the data following the official guidelines, which included maintaining a pointing offset angle of less than 0.04$^\circ$, ensuring an Earth elevation angle greater than 10$^\circ$, requiring a geomagnetic cutoff rigidity above 8$^\circ$, and excluding data collected within 300 seconds of traversing the South Atlantic Anomaly (SAA). Background estimation was performed using HEBKGMAP, MEBKGMAP and LEBKGMAP, while response files were generated through the tasks HERSPGEN , MERSPGEN and LERSPGEN.  Due to the higher background count rate in the HE band, we focused our spectral analysis with Insight-HXMT on the energy bands of 1-8.0 keV (LE) and 10-30.0 keV (ME).

\subsection{NICER}
NICER \citep{gendreau2016neutron} is an advanced X-ray telescope located on the International Space Station (ISS) that employs silicon-drift detectors, providing a sensitivity range of 0.2 to 12 keV. With an effective area exceeding 2000 cm² at 1.5 keV and 600 cm² at 6 keV,  It's time-tagging resolution is less than 300 ns. NICER has a spatial resolution of 5 arcmin in diameter with a non-imaging field of view. The background is primarily dominated by the diffuse cosmic X-ray background in the soft range. 

Between June 5 and October 7, 2018, NICER conducted a series of 86 targeted observations of MAXI J1727-203. For our analysis, we utilized the NICER data analysis threads\footnote{\url{https://heasarc.gsfc.nasa.gov/docs/nicer/analysis_threads}} based on the official HEASOFT V6.32 software, incorporating the calibration files (xti20221001). We employed the tools \textit{nicerl2} and \textit{nibackgen3C50} to extract the source and background, respectively. Additionally, we generated light curves using \textit{nicerl3-lc} in the energy ranges of 1-4 keV, 4-10 keV, and 1-10 keV.

\section{ANALYSIS AND RESULTS}
\label{RESULTS}
\subsection{Light Curves}

\begin{figure}
    \includegraphics[width=\columnwidth]{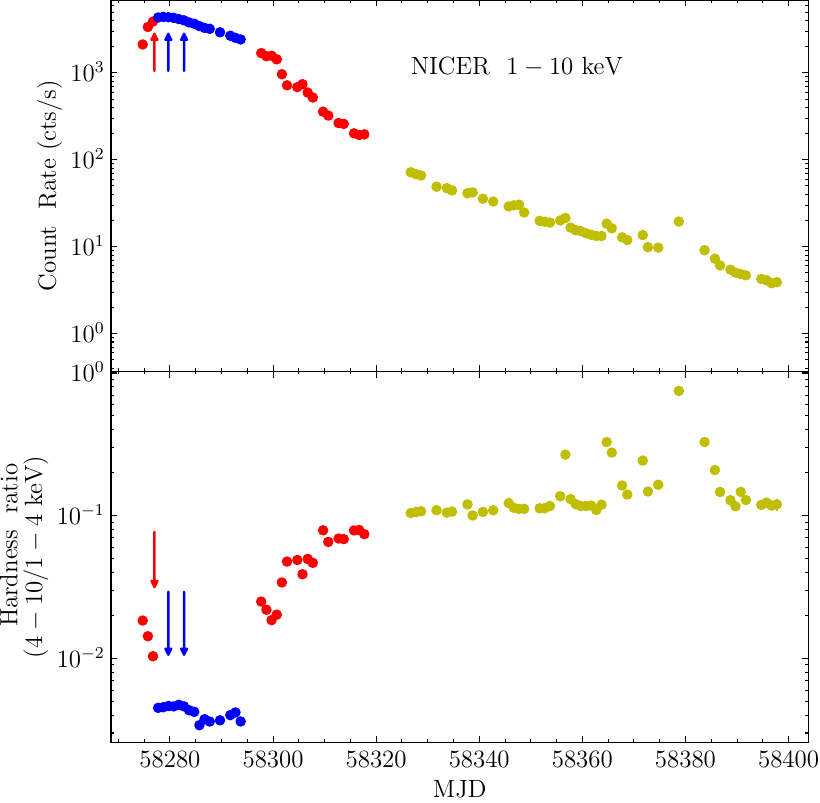}
    
    \caption{Top panel: NICER 1-10 keV light curve. All data points have been rebinned to one-day intervals, with IMS, HSS, and LHS represented in red, blue, and yellow, respectively. The three arrows indicate the times corresponding to the observations by Insight-HXMT. 
    Bottom panel: Hardness ratios derived from NICER data using (4-10 keV)/(1-4 keV). The arrows and colors have the same meanings as in the top panel. The errors for all data points are at least one order of magnitude smaller than the data points, making them not visible on the plot.}
    \label{figure1}
\end{figure}

In the top panel of Figure~\ref{figure1}, we presented the background-subtracted light curves for MAXI J1727-203, observed by NICER during the observation period.  The variation in the calculated hardness ratio is shown in the bottom panel of Figure~\ref{figure1}. All data points have been rebinned to one-day intervals, with IMS, HSS, and LHS represented in red, blue, and yellow, respectively. The time intervals for the different states are as follows: LHS from MJD 58327 to 58397, IMS from MJD 58274 to 58278 and from MJD 58298 to 58327, and HSS from MJD 58278 to 58298. Our classifications are consistent with the results of \cite{alabarta2020x}. The positions indicated by the arrows in Figure~\ref{figure1} represent the times of the three observations conducted by Insight-HXMT, while the different colors represent the states of the source during these observations.

\begin{figure}
    \includegraphics[width=\columnwidth]{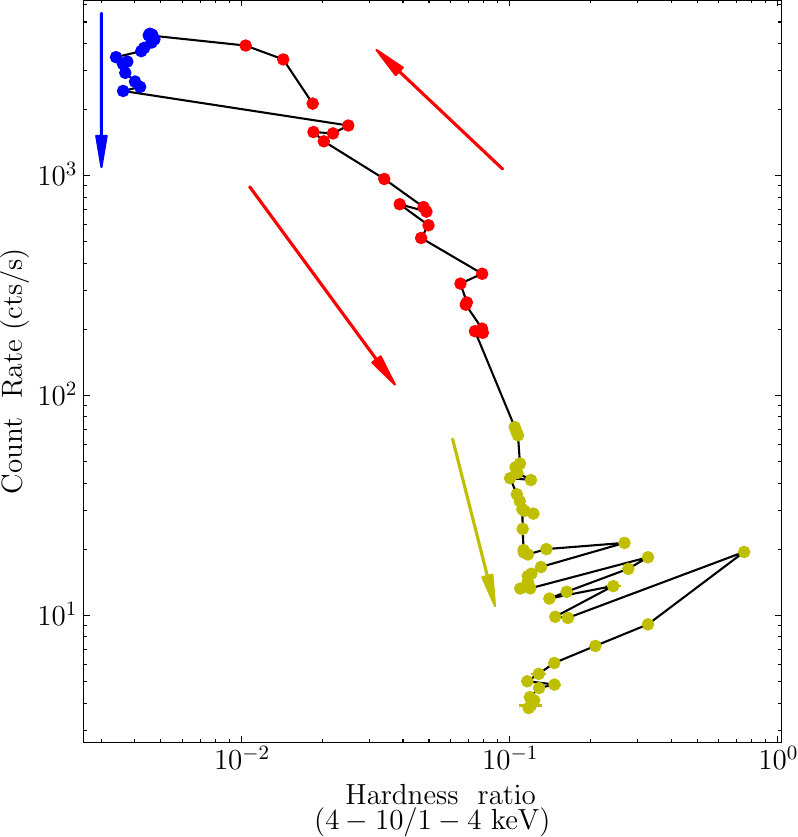}
    
    \caption{The HID of MAXI J1727-203 during NICER observations. The data points and arrow colors indicate different states, with IMS, HSS, and LHS represented in red, blue, and yellow, respectively. The direction of the arrows indicates the evolution of the outburst process.}
    \label{figure2}
\end{figure}

We present the hardness-intensity diagram (HID) of MAXI J1727-203 in Figure~\ref{figure2}, which provides a clearer view of its evolution throughout the outburst. The meanings of the colors in the figure are consistent with those in Figure~\ref{figure1}. Although we are missing the rise of the outburst, the NICER HID still displays a counter-clockwise Q-shape, signifying transitions from the IMS to the HSS at peak brightness, and from the HSS to the LHS at lower brightness levels.

\subsection{The Spectral Analysis}

During the three observations by Insight-HXMT, the first observation corresponds to the IMS, while the subsequent two observations correspond to the HSS, with the hardness ratio obtained from NICER data being less than $5\times 10^{-3}$. Therefore, we conducted spectral analysis for these two observations using LE and ME spectral data. We estimated the uncertainties of the parameters using Markov Chain Monte Carlo (MCMC) simulations, employing 8 walkers over a total of 30000
steps. The uncertainties reported in the spectral fitting correspond to the 99 percent confidence level. 

\subsubsection{Non-Relativistic Model}

\begin{figure}
    \includegraphics[width=\columnwidth]{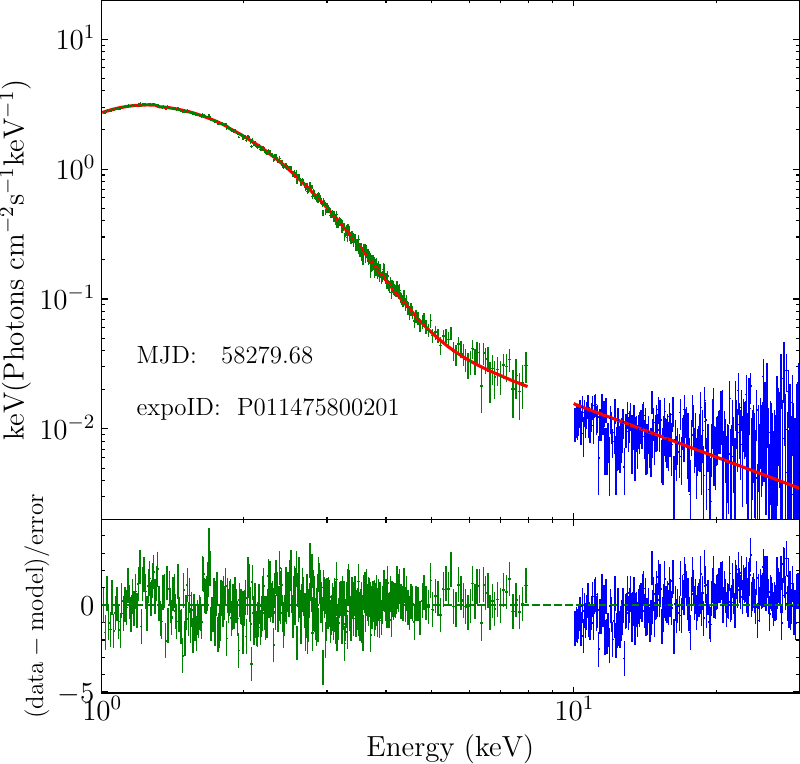}    

    \caption{The spectrum and residuals for non-relativistic model  are shown as an example. The green and blue data points correspond to LE and ME, respectively. The observation time and expoID are labeled on the plot. 
}
    \label{figure3}
\end{figure}

\begin{figure}

    \includegraphics[width=\columnwidth]{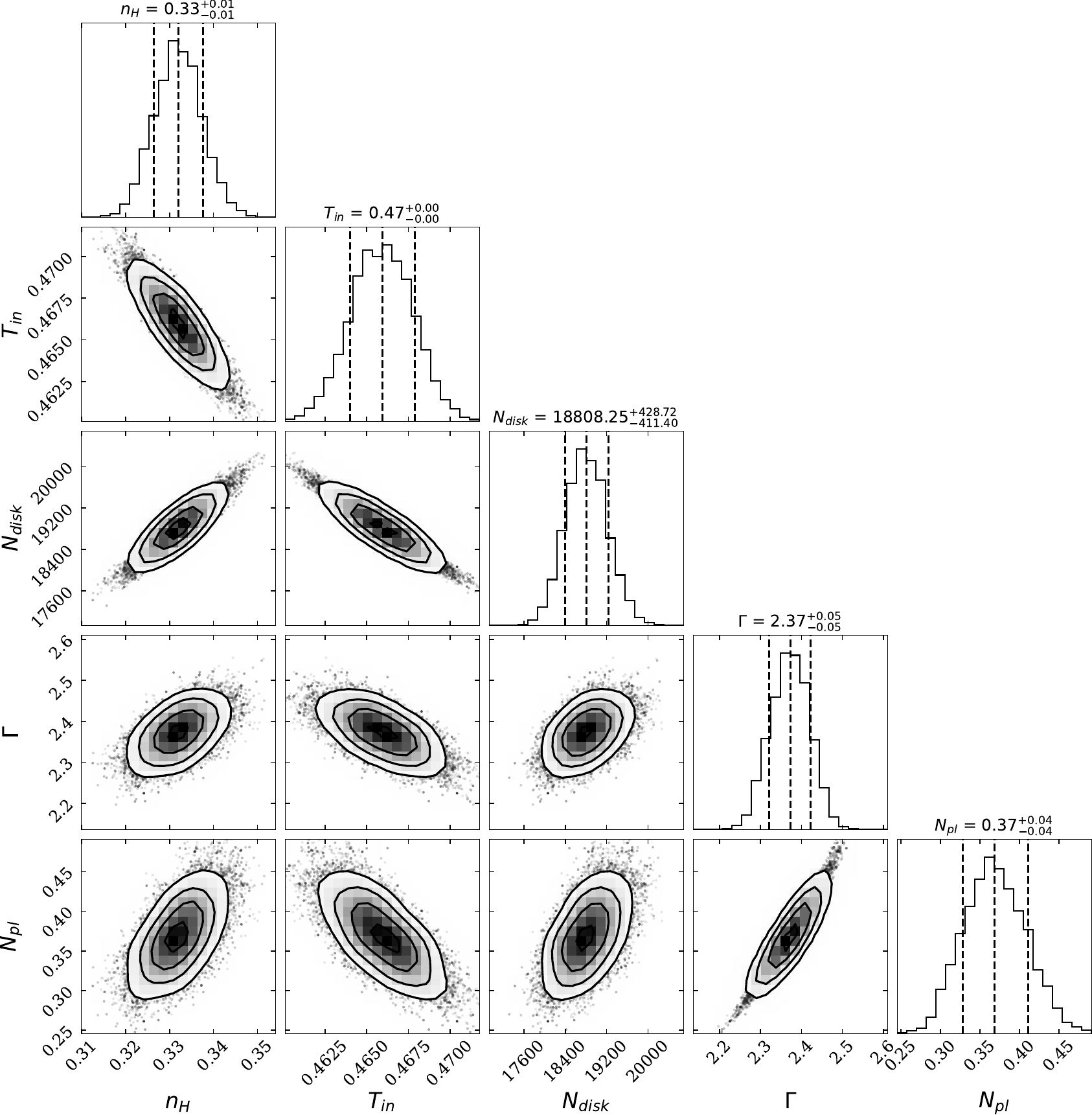}
    \caption{The corner plot of the posterior probability distributions derived from the MCMC analysis for the parameters in the non-relativistic model.
}
    \label{figure4}
\end{figure}

Firstly, the spectrum is fitted by the multicolour disk blackbody model
diskbb (Mitsuda et al. 1984; Makishima et al. 1986) and a power-law
component:
Model 1: {\tt tbabs*(diskbb +powerlaw )}. The {\tt tbabs} model is employed to account for interstellar absorption, using abundances as described by \citet{wilms2000absorption}. This model allows for the adjustment of only the equivalent hydrogen column density, $n_{\text{H}} $, in units of $ 10^{22} \, \text{atoms cm}^{-2}$. During the fitting process, we set $n_{\text{H}} $ as a free parameter. We present the fitting results in Table~\ref{tab2}, examples of the spectra and residuals in Fig~\ref{figure3}, and the probability distribution of the parameters in Fig~\ref{figure4}. The fitted disk temperature in the first two observations is approximately 0.46 keV, while in the third observation, it decreased to 0.44 keV. Concurrently, the spectral index remained around 2.3 during the first two observations but increased to about 2.8 in the third observation.

\begin{table*}
\centering
\caption{The Results of Spectral Fitting the Insight-HXMT Data for Non-Relativistic Model}
\begin{tabular}{ccccc}
 \hline\noalign{\smallskip}

Component &Parameter &ExpoID &ExpoID & ExpoID \\

 &&P011475800201 &P011475800202 &P011475800301\\
\hline
{\tt tbabs} & $n_{\text{H}}$$~\left[ \rm 10^{22}~cm^{-2} \right]$  &$ 0.33_{-0.01}^{+0.02}$&$ 0.33_{-0.03}^{+0.02}$& $ 0.39_{-0.02}^{+0.02}$ \\
{\tt diskbb}&$\rm T_{in}$$~\left[ \rm keV \right]$ & $ 0.466_{-0.006}^{+0.005}$&$ 0.467_{-0.007}^{+0.009}$& $ 0.444_{-0.008}^{+0.007}$   \\
&$\rm Norm$$~\left[ \rm 10^{4}\right]$&$ 1.9_{-0.1}^{+0.1}$&  $ 1.9_{-0.2}^{+0.1}$ &  $ 2.0_{-0.1}^{+0.2}$\\
{\tt powerlaw} &$\Gamma$& $ 2.4_{-0.1}^{+0.1}$&$ 2.4_{-0.1}^{+0.2}$  & $ 2.9_{-0.1}^{+0.1}$  \\
 &Norm& $ 0.4_{-0.1}^{+0.1}$ & $ 0.4_{-0.1}^{+0.2}$  &  $ 1.4_{-0.2}^{+0.3}$ \\
 &$\chi^2/dof$&  $ 0.77$&  $ 0.79$ &  $ 0.73$ \\
 \noalign{\smallskip}\hline
\label{tab2}
\end{tabular}
  \end{table*}

\subsubsection{Relativistic Model}
After employing a non-relativistic model, we adjusted the parameters and developed a physically realistic model by substituting the diskbb component with the kerrbb model \citep{li2005multitemperature} to determine the black hole's spin (Model 2: {\tt tbabs*(kerrbb +powerlaw )}).  

In the continuum-fitting method, parameters such as the black hole mass, source distance, and system inclination are crucial and significantly impact the results. But for the MAXI J1727-203, these parameters have not been precisely determined. \cite{wang2022multi} constrained the black hole mass to be $11.5 \, M_\odot \leq M \leq 350.5 \, M_\odot$ for a distance of $5.9 \, \text{kpc} \leq D \leq 57.3 \, \text{kpc}$. \cite{draghis2023preliminary}  obtained the inclination angle of this system to be $60_{-7}^{+10} $ degrees using the relativistic reflection method. 

\begin{figure}
    \includegraphics[width=\columnwidth]{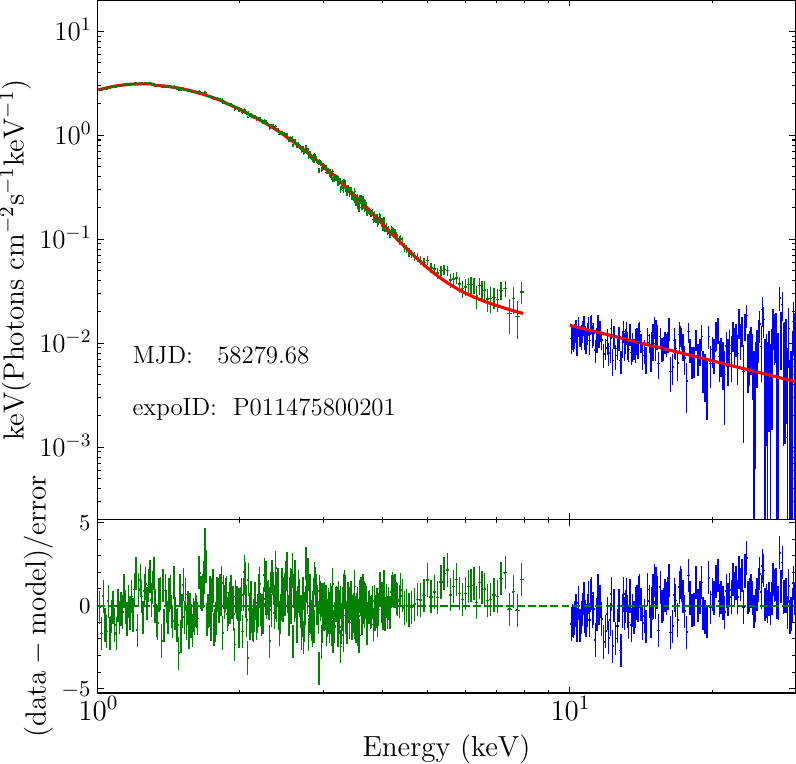}
    \caption{The spectrum and residuals for the relativistic model are shown as an example. The green and blue data points correspond to LE and ME, respectively. The observation time and expoID are labeled on the plot. 
}
    \label{figure5}
\end{figure}

\begin{figure}
    \includegraphics[width=\columnwidth]{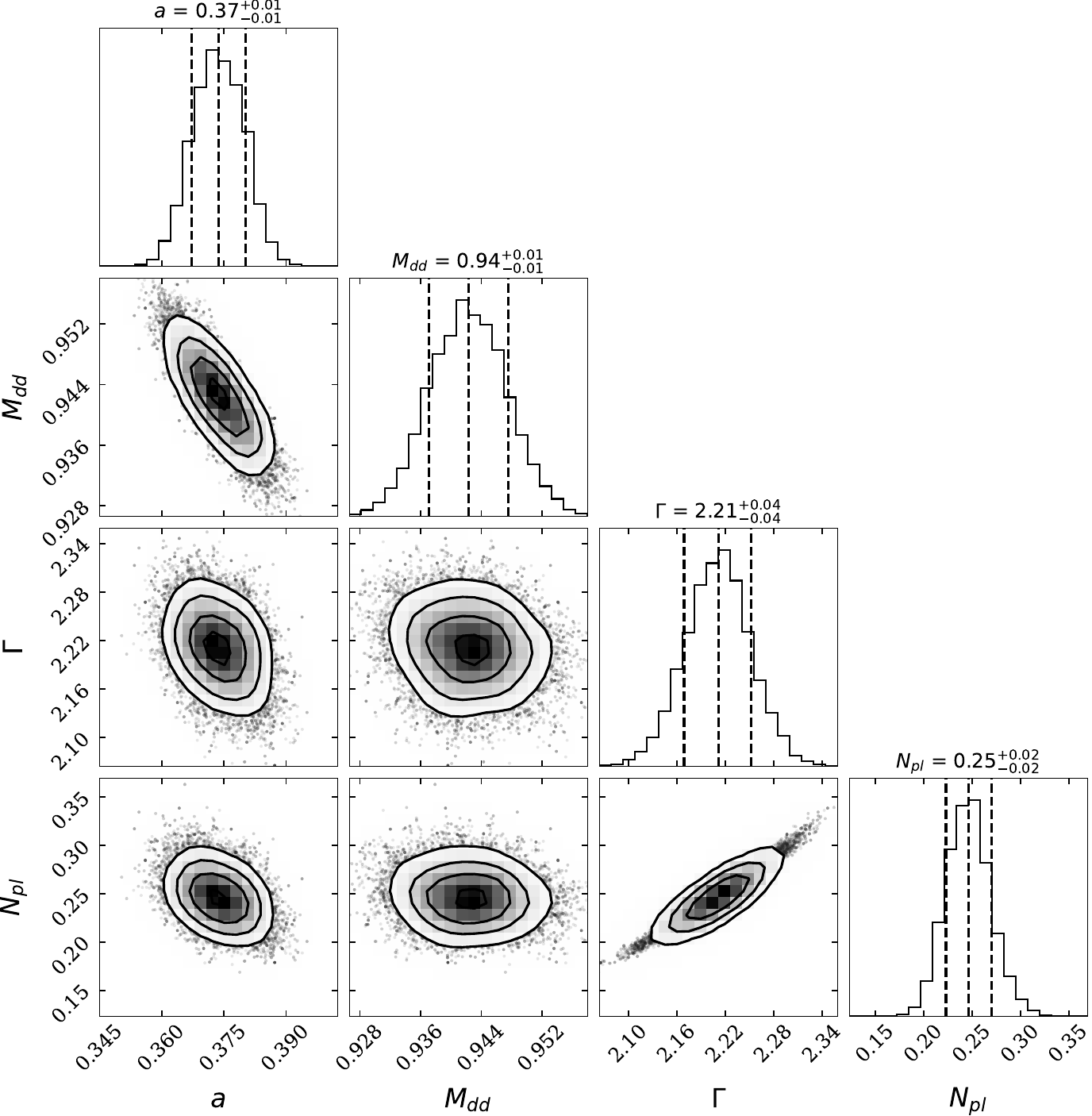}
    \caption{The corner plot of the posterior probability distributions derived from the MCMC analysis for the parameters in the relativistic model. 
}
    \label{figure6}
\end{figure}

Due to the lack of precise measurements for these parameters, we at first systematically varied the distance, mass, and inclination within the relatively broad parameter ranges previously established in our study, using a step size of 0.1. Additionally, we attempted to fit the model by allowing all three parameters to vary, and found that similar results could be obtained within the same parameter ranges. Ultimately, the best fit is achieved with a black hole mass of approximately $12 \, M_{\odot}$, a distance of around $6 \, \text{kpc}$, and an inclination angle of roughly $30^\circ$. The fitted results for the mass and distance fall within the parameter ranges reported in previous studies and are close to the lower limits of those ranges \citep{wang2022multi}. So in subsequent fittings, we fix the black hole mass at $M_{\text{BH}} = 12 \, M_\odot$, the distance at $D = 6 \, \text{kpc}$ and inclination angle at $30^\circ$, fitting the spin as free parameters. Based on the fitting results from the non-relativistic model, the equivalent hydrogen column density in {\tt tbabs} was fixed to $0.35 \times 10^{22} \, \text{cm}^{-2}$ after averaging. We also examined the impact of freeing this parameter on the fit results and found that it had a negligible effect on the final spin fitting, while the hardening factor was fixed at 1.7 \citep{shimura1995spectral}. We present the fitting results in Table~\ref{tab3}, examples of the spectra and residuals in Fig~\ref{figure5}, and the probability distribution of the parameters in Fig~\ref{figure6}. 

In addition, we also utilized the {\tt{tbabs*SIMPL*kerrbb}} model to fit the observational data (e.g. ExpoID P011475800201) to examine whether different models would significantly affect the measurement of the spin parameter. {\tt{SIMPL}} model \citep{steiner2009simple} represents the Comptonization of a seed spectrum, where a fraction of photons are scattered into a power-law distribution. The fitting results indicate that the impact of different non-thermal emission models on the spin parameter is minimal. Compared with {\tt{powerlaw}} (Model 2), the spin value changes slightly from about 0.37 to 0.34, while the mass accretion rate increases from 0.94 to 1.001. Therefore, we only used Model 2 for subsequent analysis.

\begin{table*}
\centering
\caption{The Results of Spectral Fitting the Insight-HXMT Data for Relativistic Model }
\begin{tabular}{ccccc}
 \hline\noalign{\smallskip}
Component &Parameter &ExpoID &ExpoID & ExpoID \\

 &&P011475800201 &P011475800202 &P011475800301\\
\hline
{\tt kerrbb} & $a_{\star}$  &$ 0.37_{-0.02}^{+0.02}$&$  0.38_{-0.03}^{+0.02}$& $ 0.41_{-0.01}^{+0.02}$ \\
&$\rm M_{dd}$$~\left[ \rm 10^{18}~g~s^{-1}\right]$ & $ 0.94_{-0.01}^{+0.02}$&$ 0.94_{-0.02}^{+0.03}$& $0.79_{-0.02}^{+0.02}$   \\

{\tt powerlaw} &$\Gamma$& $ 2.2_{-0.1}^{+0.1}$& $ 2.2_{-0.2}^{+0.2}$  & $ 2.6_{-0.08}^{+0.07}$  \\
 &Norm& $0.25_{-0.06}^{+0.08}$ & $ 0.24_{-0.09}^{+0.12}$  &  $0.81_{-0.12}^{+0.13}$ \\
 &$\chi^2/dof$&  $0.76$&  $ 0.79$ &  $ 0.79$ \\

\noalign{\smallskip}\hline
\label{tab3}
\end{tabular}

  \end{table*}
  
\subsubsection{Error Analysis}
When fitting the spin using the continuum-fitting method, parameters such as distance, inclination, and mass have a significant impact on the final fitting results \citep{mcclintock2015black}. In line with previous work, we apply the Monte Carlo (MC) method to conduct error analysis \citep{gou2010spin,zhao2020confirming,wu2023moderate,guan2024estimating}. 

Due to the lack of precise measurements for the distance, inclination, and mass of MAXI J1727-203, which are only roughly estimated, our fitting results indicate the best fit around $(D, i, M) \approx (6 \text{kpc}, 30^\circ, 12M_{\odot})$. Thus, we set over 3000 sets of parameters within the ranges of $D \sim (5.9\, \text{kpc}, 7\, \text{kpc}) $, $i \sim (24^\circ, 35^\circ)$, and $M \sim (10\, M_{\odot}, 14\, M_{\odot}) $. We assume that these parameters are independent and follow a uniform distribution. These parameters were evenly distributed to investigate the influence of different factors. Subsequently, we employed a relativistic model to fit the energy spectra from the three exposures, enabling us to determine the distribution of the spin. 

\begin{figure*}
    \includegraphics[width=\textwidth]{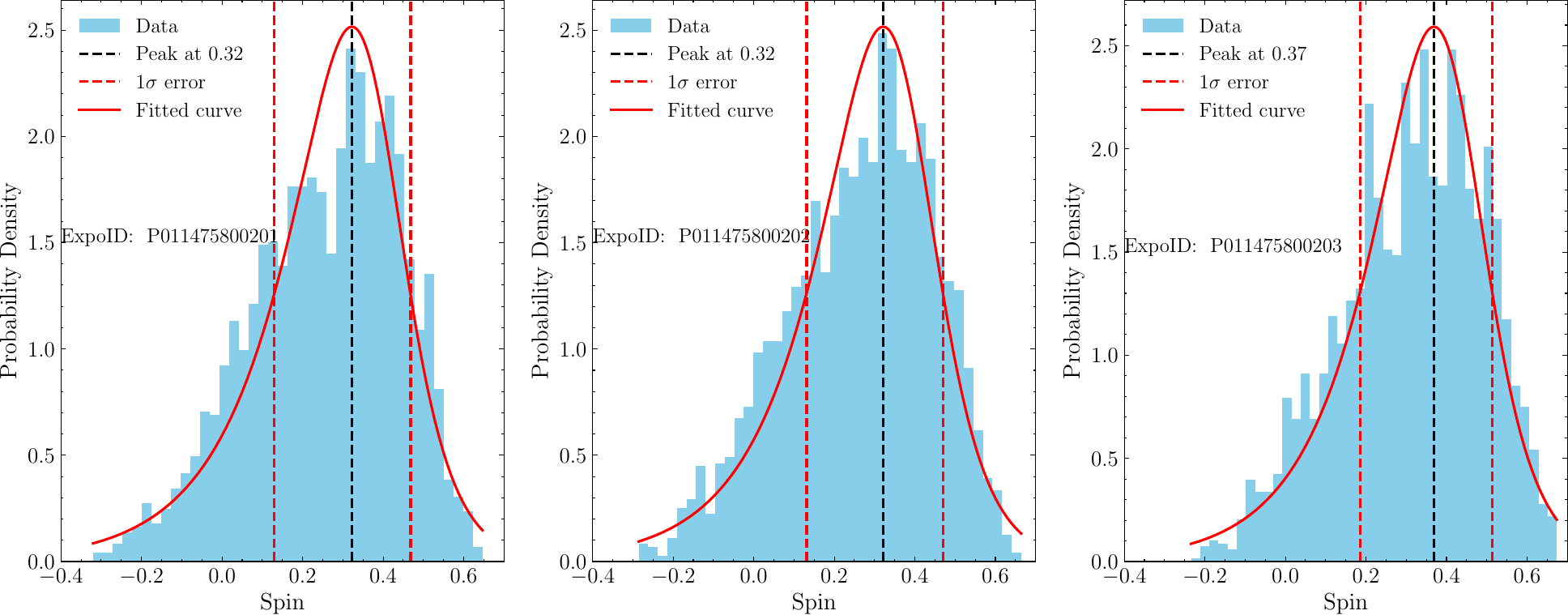}

    \caption{The distribution of $a_{\star}$
  from the Monte Carlo method for the fitting results of the three exposures. The red dash lines represent the 1$\sigma$ error, the black dashed lines indicate the best-fitting value, and the red line shows the generalized logistic distribution fitting. The corresponding expoID is labeled on the plot.}
    \label{figure7}
\end{figure*}

\begin{figure}
    \includegraphics[width=\columnwidth]{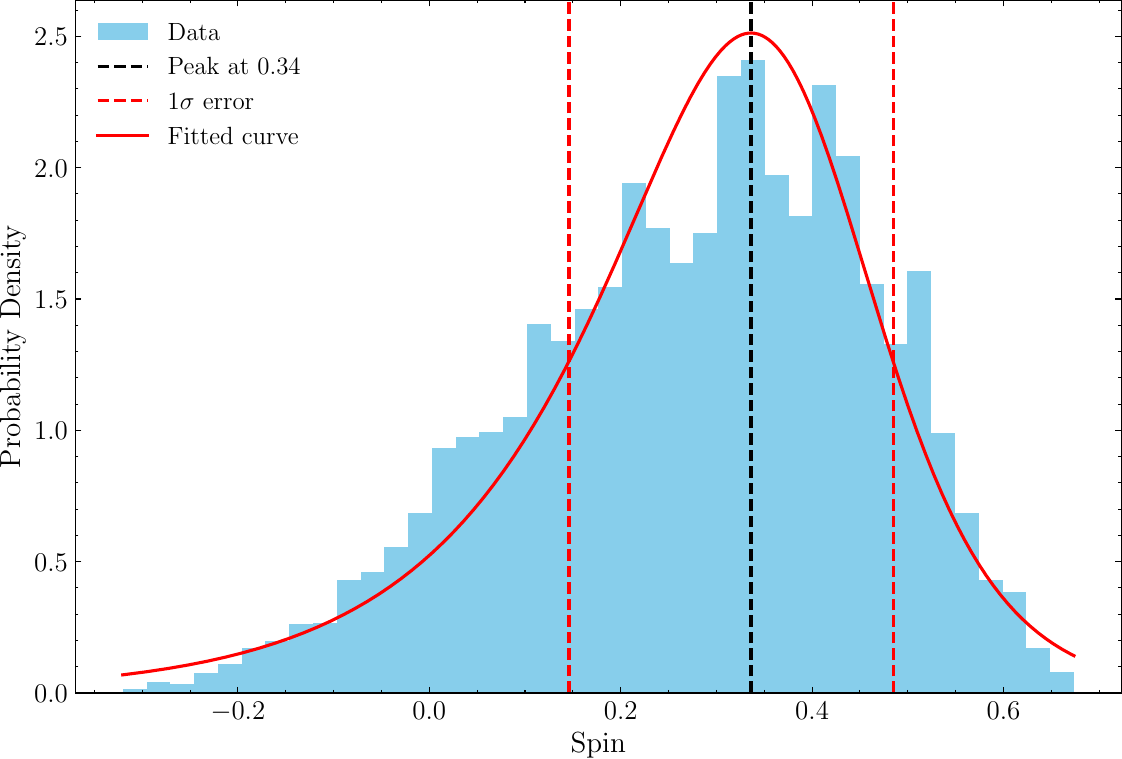}

    \caption{The summed distribution of 
$a_{\star}$ from the Monte Carlo method for the fitting results of the three exposures, with a total of 9000 data points. The red dash lines represent the 1$\sigma$ error, the black dashed lines indicate the best-fitting value, and the red line shows the generalized logistic distribution fitting. 
}
    \label{figure8}
\end{figure}

During the fitting process, we found that many parameters did not yield satisfactory results, and in several parameter spaces, the fitted spin values were retrograde (for a detailed discussion, see the Sections \ref{DISCUSSION}). The distribution of spin for each spectrum is displayed in Fig~\ref{figure7}, and the combined histogram is presented in Fig~\ref{figure8}. 
We used the generalized logistic distribution to fit the spin distribution we obtained. Its probability density function is: 
\begin{equation}
 f(x,c)=c \frac{\exp(-x)}{(1+\exp(-x))^{c+1}}, 
\end{equation}
where $c$ is the shape parameters. In Fig~\ref{figure7}, we present the spin distribution obtained from the three observations and the fitted curves. The red solid line represents the fitted probability density distribution, while the position of the black dashed line corresponds to the peak of the probability density distribution. The red dashed line marks the position corresponding to $1\sigma$. The corresponding ExpoIDs are labeled on the plot.
Fig~\ref{figure8} shows the combined histogram and the corresponding fitted curve. The markings in the figure are consistent with those in In Fig~\ref{figure7}. The final spin value we obtained is $a=0.34_{-0.19}^{+0.15}$ ($1\sigma$). 

\section{DISCUSSION} \label{DISCUSSION}
\subsection{Effect of Different Parameter Configurations}

\begin{figure*}
    \includegraphics[width=\textwidth]{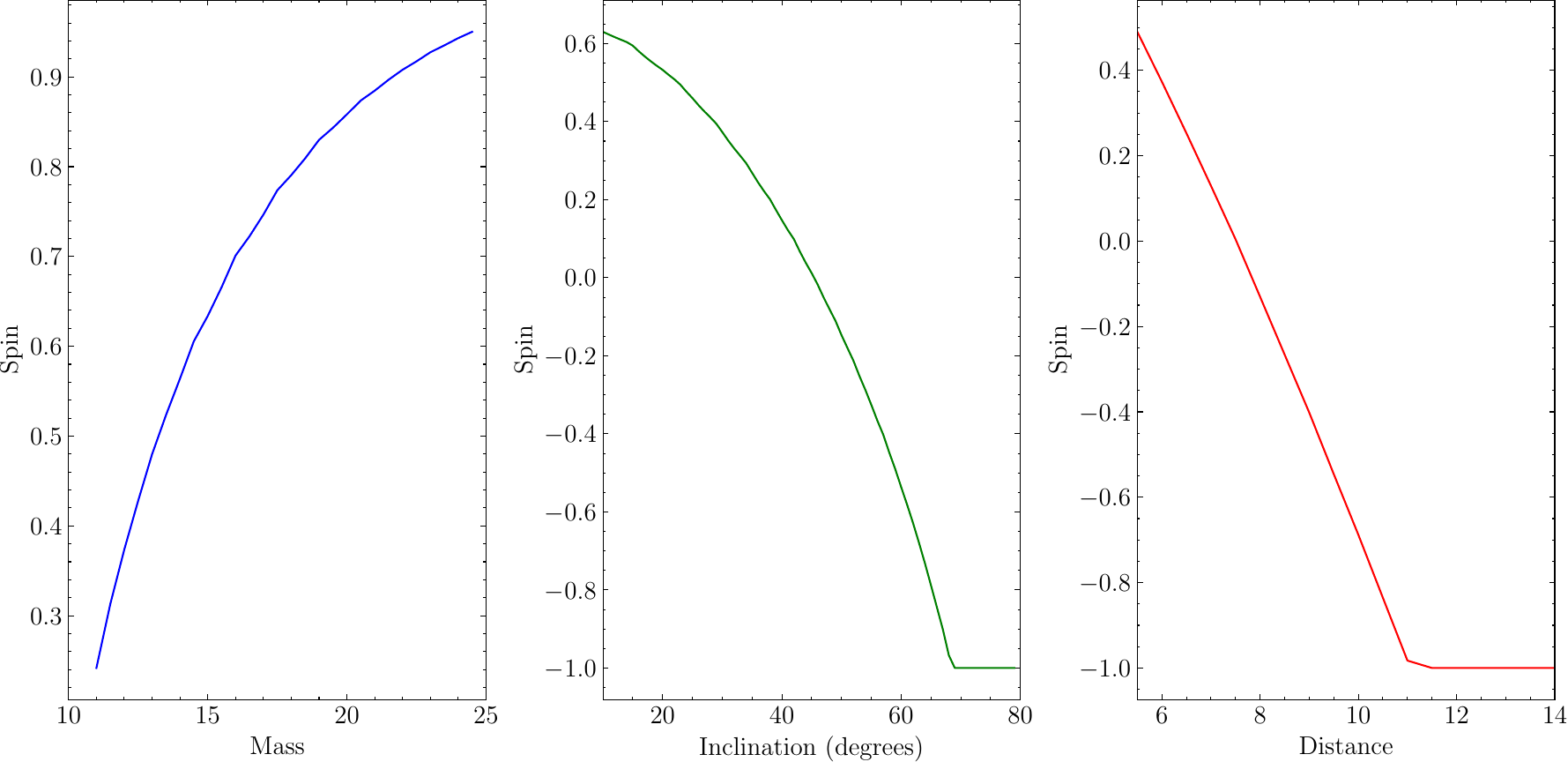}
    \caption{Correlation plots illustrating the impact of varying $M$, $i$, and  $D$ on measuring the spin.
}
    \label{figure9}
\end{figure*}

As we mentioned above, after traversing the entire parameter space, we obtained a good fit to the spectrum within a relatively narrow range. The variations in mass, distance, and inclination angle can affect the measurement of the black hole spin. For this source, these parameters have significant uncertainties; therefore, we need to examine the impact of different parameter values on the results of the spin measurement. In Fig~\ref{figure9}, we present the relationship between the parameter and the spin over a broader range. From the figure, it can be observed that the spin value gradually increases as the black hole mass increases. However, as the inclination angle and distance increase, the spin decreases sharply, leading to a retrograde black hole. 

Since most the black hole spins obtained using the continuum-fitting method are positive \citep{steiner2011spin,steiner2016spin,chen2016spin,zhao2021estimating,feng2023using,wu2023moderate,guan2024estimating,sai2024revisiting,chen2024spin}, and there has been no clear observational evidence for the existence of retrograde black holes, this suggests that the results of a retrograde black hole must be interpreted with caution.

\cite{morningstar2014spin} initially reported a retrograde spin for the black hole in Nova Muscae 1991, with a spin parameter of $a=-0.25_{-0.64}^{+0.05}$ (90$\%$ confidence level). However, they noted that this measurement could be better constrained if the distance to the binary and the mass of the black hole were more accurately determined. Subsequently, with more accurate estimates of the black hole mass, the orbital inclination angle of the system and the distance, \cite{chen2016spin} obtained a spin parameter of $a=0.63_{-0.19}^{+0.16}$ ($1\sigma$). 

In the study of MAXI J1659-152, \cite{feng2022}, based on the more accurate distance, mass, and inclination information provided by \cite{torres2021delimiting} through spectroscopy, with $i = 70^\circ - 80^\circ$, $D = 6 \pm 2$ kpc, and $M = 4.9 - 5.7 M_{\odot}$, investigated the spin of this source using the continuum-fitting method.
They found that the spin parameter was constrained to $-1 < a \lesssim 0.44$ (90\% confidence), with an inverse correlation to the inclination angle, then for the young age of this system, an extreme retrograde spin remains possible.

\begin{figure}
    \includegraphics[width=0.5\textwidth]{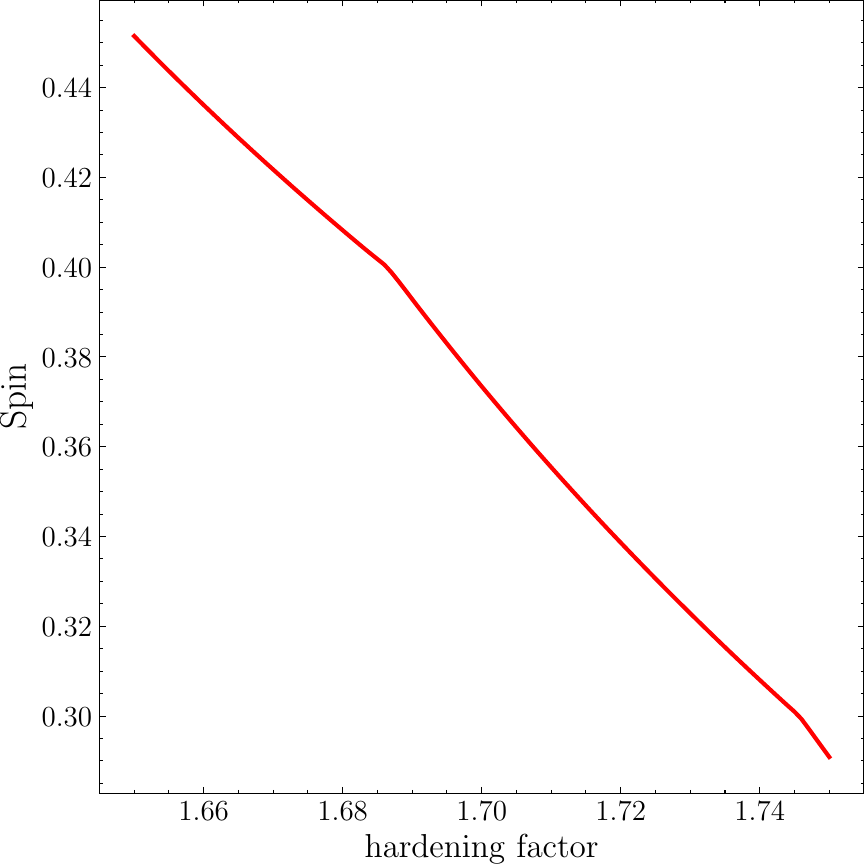}
    \caption{Correlation plots illustrating the impact of varying hardening factor on measuring the spin.
}
    \label{hf}
\end{figure}

In the {\tt kerrbb} model, a strong correlation exists between the accretion rate (or luminosity) and the spin parameter. As the spectral hardening factor can vary with luminosity \citep{davis2006testing}, it is also inversely correlated with the spin parameter, as shown in Figure 4 of \cite{salvesen2021black}. Previously, we fixed the hardening factor at 1.7, so we also treat it as a free parameter to investigate its impact on the measured spin values. We chose the first observation as an example to perform the spectral fittings again, allowing the hardening factor to vary between 1.5 and 1.8, during which the spin parameter changed from about 0.65 to 0.2. Subsequently, we generated 100 evenly spaced values for the hardening factor in the range of 1.65 to 1.75 and performed fittings to derive the relationship between the spin parameter and the hardening factor which is presented in Figure~\ref{hf}. It can be seen that there is an inverse correlation between the spin parameter and the hardening, consistent with previous studies \citep{salvesen2021black,wu2023moderate,guan2024estimating}.

\subsection{Re-fitting of NuSTAR spectra in LHS}

\cite{draghis2023systematic} applied the reflection method based on the relxill family of models to estimate the spin of MAXI J1727-203 using a NuSTAR observation taken during the phase when the source had just entered the LHS.  However they found an extreme black hole spin of $0.986_{-0.159}^{+0.012}$  and a high inclination of $64_{-7}^{+10}$ degrees. The spin value obtained from the reflection fit is significantly higher than the result based on the continuum model in this work. To resolve the difference of the spin measurement for two methods, we also analyzed the NuSTAR spectra during the low hard state (LHS) to constrain the possible reflection components. Here, we focused on the NuSTAR observation of MAXI J1727-203 with ObsID 90401329002 as the example, a detailed description of the data processing procedure is also provided in the Appendix~\ref{secapp}. 

\begin{figure}
    \includegraphics[width=\columnwidth]{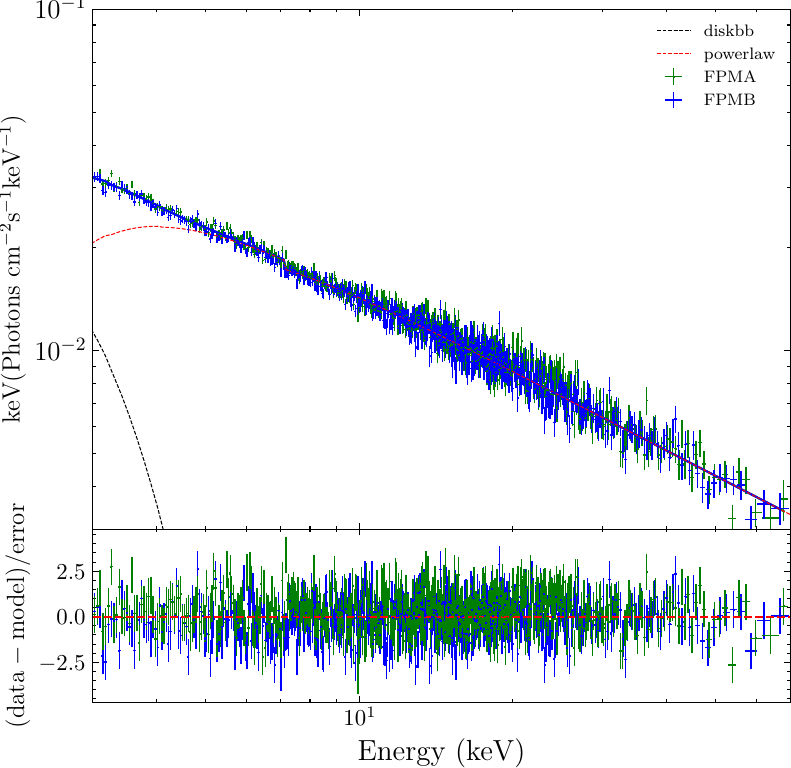}
    \caption{The top panel shows the spectrum of MAXI J1727-203, with the NuSTAR FPMA spectrum in green and the FPMB spectrum in blue. The reported best-fit model {\tt tbabs*(diskbb +powerlaw )} is represented by the blue solid lines, while the contribution of the {\tt diskbb} component is shown by the black dashed line. The red dashed line represents the {\tt powerlaw} component. The bottom panel displays residuals. The red dashed line marks the position where the residuals are zero. }
    \label{figure10}
\end{figure}

We used the {\tt tbabs*(diskbb +powerlaw )}  model to fit the energy spectra of FPMA and FPMB. We allowed the hydrogen column density of \textit{tbabs} to vary, resulting in $\chi^2/\nu = 1749.07/1772 = 0.987$. The fitting result for the hydrogen column density is $(3.948 \pm 0.466) \times 10^{22} \text{cm}^{-2}$, in the \textit{diskbb} model, the inner disk temperature ($T_{\text{in}}$) is $0.500 \pm 0.013$ keV, and the normalization ($\textit{diskbb}_{norm}$) is $516.209 \pm 141.615$. In the \textit{powerlaw} model, the photon index ($\Gamma$) is $1.77 \pm 0.007$. We present the fitting results in Fig~\ref{figure10}, we can observe that there are no significant reflection features in the residuals. Therefore, we believe that the previous studies, when fitting with the \textit{tbabs} $\times$ (\textit{diskbb} + \textit{powerlaw}) model, may have improperly fitted the parameters, leading to the appearance of the iron line feature in the residuals \citep{draghis2023systematic}. This likely caused the large discrepancy between the spin values obtained using the reflection model and those we derived. However, since the detailed fitting parameters of their results were not presented, we are unable to make a direct comparison. 
In addition, \cite{negoro2018maxi} found the disk component below 3 keV, while the energy range used in the NuSTAR data fitting is above 3 keV, which could also potentially cause iron line issues with the fitting results. Subsequently, we will conduct a detailed spectral analysis of the NICER observations of MAXI J1727-203 during its outburst, hoping to identify possible prominent reflection features. 

It is worth noting that the spin values of the black holes in X-ray binaries obtained from fitting using the reflection method are generally larger than those derived from continuum-fitting \citep{Reynolds2021}. In the spin measurement of MAXI J1348-630, \cite{song2023spin} used the reflection model to fit the Insight-HXMT data and constrained the spin parameter to be $0.82_{-0.03}^{+0.04}$ ( 90 percent confidence). \cite{draghis2024systematically} measured the spin of MAXI J1348-630 based on relativistic reflection, obtaining a spin value of $0.977_{-0.055}^{+0.017}$ and an inclination angle of $52_{-11}^{+8}$ degrees. This result also shows a significant difference compared to the spin value of $0.78_{-0.04}^{+0.04}$ and inclination angle of $29.2_{-0.5}^{+0.3}$ degrees based on the reflection model using NuSTAR data reported by \cite{jia2022detailed}. \cite{jia2022detailed} fixed the emissivity indices ($q_1 = q_2 = 3$), meaning their spin measurement represents a lower limit. They also did not account for narrow absorption features in the spectra. \cite{draghis2024systematically} results show higher spin values, as expected when allowing the emissivity parameters to vary freely. Additionally, the inclusion of Gaussian absorption features increases the inclination angle, leading to a higher measurement. Recently, \cite{wu2023moderate} obtained a spin of $0.42^{+0.13}_{-0.50}$ using the continuum-fitting method. \cite{guan2024estimating} estimated the spin of MAXI J1348-630 using the continuum-fitting method, obtaining a value of $0.79 \pm 0.13$. They suggest that the accuracy of spectral fitting, particularly of the continuum spectrum, may affect the reflection component and spin measurements.


For the other black hole candidate 4U 1543-47, a similar situation also occurred. \cite{yang2024measuring} used observations from Insight-HXMT and employed the X-ray reflection fitting method to analyze the spectral data and measure the spin, and estimated the spin parameter of the black hole to be $0.902_{-0.053}^{+0.054}$ (90 percent confidence). \cite{chen2024spin} determined the spin of 4U 1543-47 using the thermal continuum-fitting method, which is sensitive to parameters such as black hole mass, distance, and inclination. Adopting their preferred values of $M = 9.4 \pm 1 \, M_{\odot}$, $D = 7.5 \pm 0.5 \, \text{kpc}$, and $i = 36.3^{+5.3}_{-3.4} \, \text{degrees}$, they obtained a moderate spin of $a = 0.46 \pm 0.12$.

\section{CONCLUSIONS}
\label{conclusion}
In this work, we present the BH spin results for MAXI J727-203 based on Insight-HXMT observations during the HSS using the continuum-fitting method. Since the spin of the black hole is highly sensitive to the measurements of the disk inclination, black hole mass, and distance for the binary system, the absence of dynamical measurements results leads to significant uncertainties in these parameters, making it challenging to accurately assess the spin. Based on previous rough estimates of these parameters, we explored the entire parameter space, and found that within the consistent value ranges, we achieved good fitting results around $(D, i, M) \approx (6\ \text{kpc}, 30^\circ, 12M_{\odot})$ using the relativistic model. Furthermore, due to the considerable uncertainty in the input parameters $D$, $M$, and $i$, we tested the influence of these parameters using Monte Carlo methods. Over 3000 sets of parameters were randomly sampled within the ranges $D \sim (5.9 \, \text{kpc}, 7 \, \text{kpc})$, $i \sim (24^\circ, 35^\circ)$, and $M \sim (10 M_{\odot}, 14 M_{\odot})$ to fit the spectra. By combining the spin distributions from these fits, we obtained a spin value of $a=0.34_{-0.19}^{+0.15}$ (1$\sigma$). The hardening factor also affects the measurement of $a$ in the continuum-fitting model, and there exists an anti-corelation between the spin parameter and hardening factor. Additionally, we processed the NuSTAR observations in the LHS. We found that using the {\tt tbabs*(diskbb +powerlaw )}  model provided a good fit, with no significant iron line features observed in the residuals. We believe this is the reason for the discrepancy between the high spin values obtained from previous reflection model fits and our continuum-fitting results.
\section*{Acknowledgements}
We are grateful to the referee for the suggestions to improve the manuscript.
This work is supported by the National Key Research and Development Program of China (Grants No. 2021YFA0718503 and 2023YFA1607901), the NSFC (12133007), the Youth Program of Natural Science Foundation of Hubei Province (2024AFB386) and the Postdoctoral Fellowship Program (Grade C) of China Postdoctoral Science Foundation (Grant No. GZC20241282). This work has made use of data from the \textit{Insight-}HXMT mission, a project funded by the China National Space Administration (CNSA) and the Chinese Academy of Sciences (CAS).


\bibliography{sample631}{}
\bibliographystyle{aasjournal}

\clearpage
\appendix
\renewcommand*\thetable{\Alph{section}.\arabic{table}}
\renewcommand*\thefigure{\Alph{section}\arabic{figure}}
\section{NuSTAR OBSERVATION}
\label{secapp}
\setcounter{figure}{0}
MAXI J1727-203 was observed by NuSTAR on July 26, 2018 (ObsID: 90401329002). In accordance with standard procedures \footnote{\url{https://heasarc.gsfc.nasa.gov/docs/nustar/analysis/}}, we processed the NuSTAR data using HEAsoft v6.32 with calibration files v20230816. To extract the source spectra, we used a circular region centered on the source position with a radius of 120\textquotesingle\textquotesingle, and background spectra were extracted using the same radius. Standard data processing tools were employed to generate level 2 (\textit{nupipeline}) and level 3 (\textit{nuproducts}) products. All spectra were grouped to ensure a minimum of 25 counts per energy bin, and $\chi^2$ statistics were applied. 

\begin{figure}[h]
\centering
    \includegraphics[width=0.4\columnwidth]{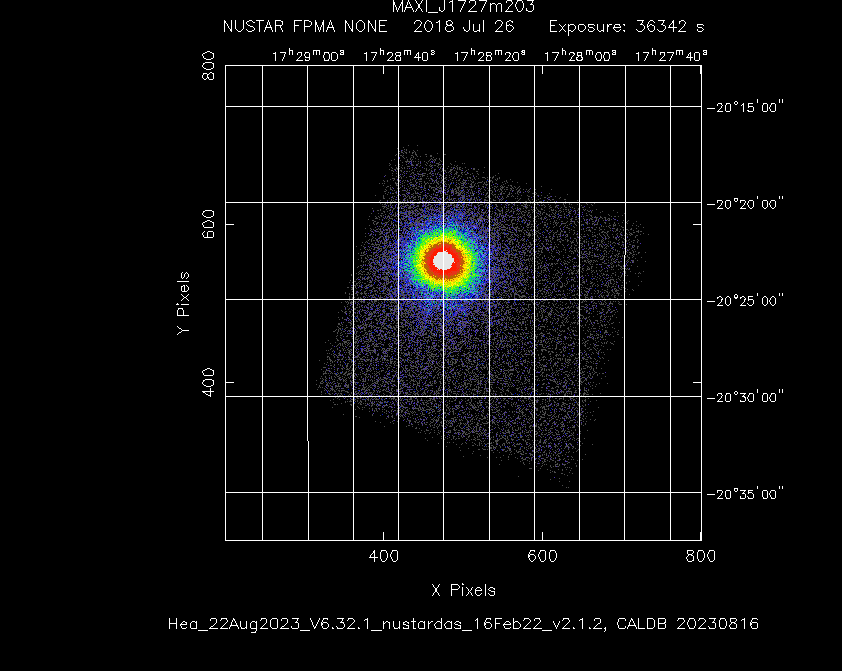}
    \includegraphics[width=0.4\columnwidth]{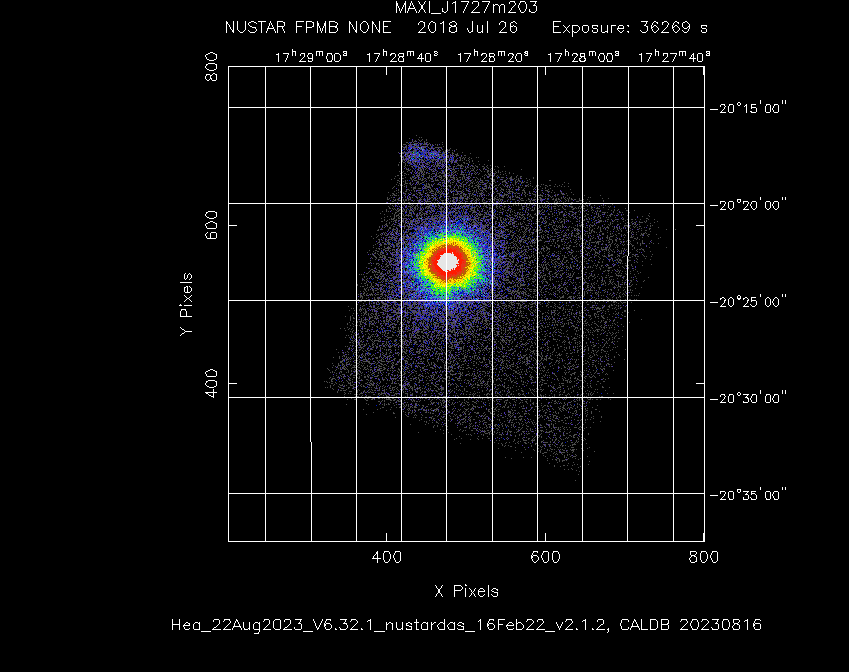}

    \caption{The position of MAXI J1727-203 in the NuSTAR field of view is shown. The left panel displays the FPMA, while the right panel shows the FPMB. The software version and CALDB version used for data processing are indicated on the plot.
}
    \label{app1}
\end{figure}

\end{document}